\documentclass[letterpaper]{article} % DO NOT CHANGE THIS
\usepackage[preprint]{aaai2027}      % DO NOT CHANGE THIS

\usepackage{times}                   % DO NOT CHANGE THIS
\usepackage{helvet}                  % DO NOT CHANGE THIS
\usepackage{courier}                 % DO NOT CHANGE THIS
\usepackage[hyphens]{url}            % DO NOT CHANGE THIS
\usepackage{graphicx}                % DO NOT CHANGE THIS
\usepackage{natbib}                  % DO NOT CHANGE THIS
\usepackage{caption}                 % DO NOT ADD OPTIONS
\usepackage{amsmath}
\usepackage{amsfonts}
\usepackage{amssymb}
\usepackage{amsthm}
\usepackage{bm}
\usepackage{nicefrac}

\usepackage{array}
\usepackage{booktabs}
\usepackage{multirow}
\usepackage{tabularx}
\usepackage{colortbl}

\usepackage{subcaption}
\usepackage{float}

\usepackage[most]{tcolorbox}
\usepackage{enumitem}
\usepackage{pifont}
\usepackage{multicol}

\usepackage{soul}
\usepackage{microtype}
\usepackage[scaled=1]{inconsolata}
\usepackage[switch]{lineno}
\usepackage{xr}
\definecolor{RecastBlueLight}{HTML}{E6F0F5}
\definecolor{RecastGain}{HTML}{117A65}
\definecolor{RecastDegradation}{HTML}{C44237}

\usepackage{algorithm}
\usepackage{algpseudocode}

\usepackage{newfloat}
\usepackage{listings}
\DeclareCaptionStyle{ruled}{labelfont=normalfont,labelsep=colon,strut=off} % DO NOT CHANGE THIS
\floatstyle{ruled}
\newfloat{listing}{tb}{lst}{}
\floatname{listing}{Listing}

\newcommand{\PP}[1]{\vspace{0em}\noindent\textbf{#1}.}

\definecolor{ASRChangeColor}{HTML}{C44237}

\usepackage{booktabs}

\title{Before Agents Speak: Pre-hoc Failure Risk Inference in Multi-Agent Systems}

\author{
Shi Lin\equalcontrib\textsuperscript{\rm 1},
Chenpei Wang\equalcontrib\textsuperscript{\rm 1},
Peng Qian\corresponding\textsuperscript{\rm 1},\\
Dezhang Kong\textsuperscript{\rm 2},
Minghao Li\textsuperscript{\rm 2},
Yufeng Li\textsuperscript{\rm 2},
Xun Wang\textsuperscript{\rm 1}
}

\affiliations{
\textsuperscript{\rm 1}Zhejiang Gongshang University\\
\textsuperscript{\rm 2}Zhejiang University\\
\{linshizjgsu,lanshuitear,messi.qp711,mhaoli9860,liyufenggood\}@gmail.com,\\
kdz@zju.edu.cn, wx@zjgsu.edu.cn
}

\begin{document}

\maketitle

\begin{abstract}
LLM-based multi-agent systems (MAS) have exhibited remarkable capabilities in collaborative reasoning and decision-making, yet their interconnected communications introduce new systemic risk: {localized hallucinations can propagate along agent communication chain, amplify through interactions, and ultimately trigger cascading failures.} Existing countermeasures predominantly follow a \textit{post-hoc} paradigm, identifying failures only after unsafe behaviors emerge, by which time harmful effects may have already spread throughout the agent network. To tackle this problem, we investigate a complementary \emph{pre-hoc} approach and propose HalluProp, a \underline{\textbf{Prop}}agation-aware \underline{\textbf{Hallu}}cination inference framework that estimates individual agent failures and emergent system-level hallucination risks before inter-agent interaction. First, we model intrinsic hallucination risks by identifying fine-grained semantic misalignment between agent roles and task queries. We then characterize inter-agent risk propagation by modeling both semantic influence and communication topology. Finally, we integrate these two risks via a differentiable Noisy-OR inference mechanism to derive a systemic diagnosis. Extensive experiments show that HalluProp accurately localizes faulty agents, achieving an average AUROC of $84.6\%$, while enabling sub-second diagnosis with over $65\times$ speedup over post-hoc methods. By facilitating early intervention through upstream screening, HalluProp effectively complements post-hoc methods, highlighting the potential of pre-hoc risk inference for building more reliable multi-agent systems.
\end{abstract}

\section{Introduction}

The rapid advancement of large language models (LLMs) has enabled the emergence of autonomous agents with multi-step reasoning and decision-making capabilities~\cite{yang2025direct}. Intuitively, LLM-based agents have made substantial progress across diverse domains, such as mathematics~\cite{du2023improving} and code generation~\cite{hong2023metagpt,yang2024swe}. In multi-agent settings, role specialization and collaborative communication enable agents to collectively solve complex problems beyond the reach of a single agent~\cite{qian2025scaling}.

Notwithstanding these unprecedented advances, MAS introduces unique and underexplored security challenges~\cite{kong2025survey,lin2024llms,ying2025pushing,kong2026web}. Among them, hallucination poses a particularly severe threat~\cite{zhang2024psysafe,huang2025survey,kim2025towards}. Unlike in single-agent scenarios, where hallucinations typically manifest as isolated factual errors, hallucinations in multi-agent systems can propagate through inter-agent communication and become progressively reinforced. In such a \textit{cascading hallucination} setting, incorrect information produced by one agent is adopted, echoed, and amplified by others, eventually distorting the collective result of the entire system. Such cascading effects can be especially harmful in high-stakes applications, e.g., healthcare, legal reasoning, and financial business analysis~\cite{chen2025medsentry}.

\begin{figure}
    \centering
    \includegraphics[width=1\linewidth]{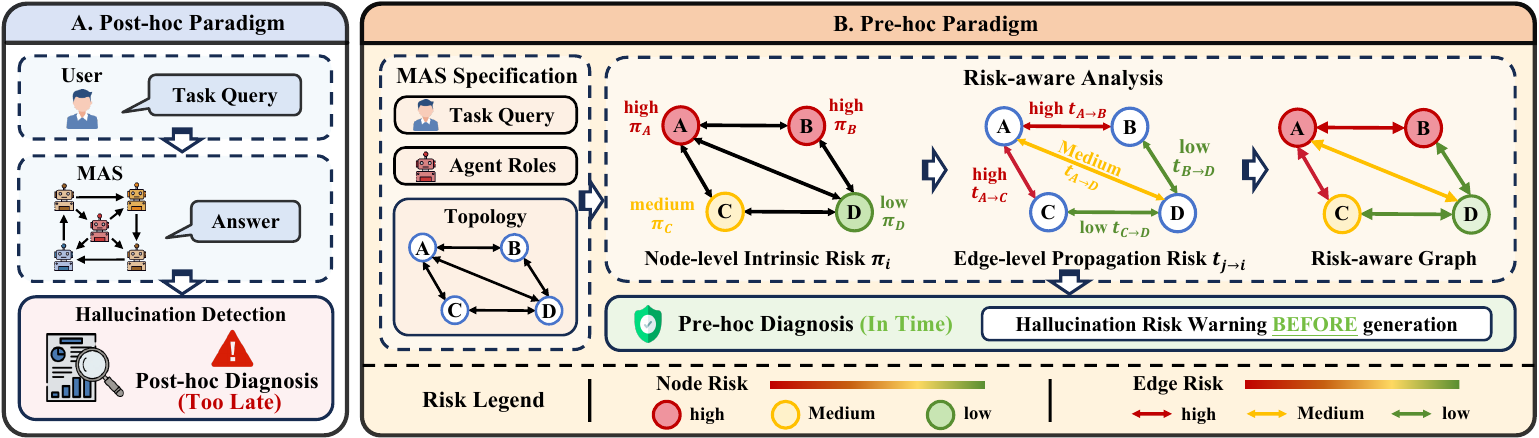}
    \vspace{-1.1em}
    \caption{Post-hoc methods detect hallucinations from observed responses and interactions, whereas pre-hoc methods predict agent failure risks before inter-agent interaction, enabling early intervention against cascading hallucinations.}
    \vspace{-0.6em}
    \label{fig1}
\end{figure}

However, hallucination diagnosis in MAS remains predominantly in a post-generation manner. Existing efforts~\cite{manakul2023selfcheckgpt,min2023factscore,tang2024minicheck,li2023halueval,zhualleviating} mostly assess hallucinations from each agent's responses. These methods are ill-suited for MAS, as they fail to capture the dynamic risks arising from inter-agent dependencies and communication topologies. More recent approaches analyze agent interactions to attribute system failures to the relevant agents~\cite{shen2025understanding,zhou2026guardian}. Although these methods identify manifested failures, they remain reactive by relying on observed agent behaviors and outputs. Furthermore, attributing failures within lengthy and entangled interaction trajectories makes timely intervention challenging. This raises an important yet unexplored question: \textit{can we infer agent-level failure risks before inter-agent interaction to enable proactive risk localization and intervention?}

% \textit{can we identify the agent contributing to the risk before it propagates to enable proactive attribution and intervention?}
To address this question, we investigate pre-hoc agent failure risk estimation before inter-agent interactions unfold. We observe that agent failures stem from two complementary sources: intrinsic risks within individual agents and interaction-induced risks arising from communication with others. Specifically, intrinsic risk arises from factors such as underspecified roles, conflicting objectives, or adversarial manipulation, whereas interaction-induced risk emerges from semantic influence and communication structures among agents. Integrating these risk sources enables system-level inference of agent-specific failure risks. Such early estimation allows proactive risk screening and localization of potential cascading failures before harmful behaviors propagate.

To this end, we propose HalluProp, a \underline{\textbf{Prop}}agation-aware \underline{\textbf{Hallu}}cination inference framework tailored for multi-agent systems, which consists of three main components. \textit{(i) Intra-agent intrinsic modeling.} First, the hallucination risk of each agent is quantified by characterizing the fine-grained semantic misalignment between its role specifications and user queries, capturing the agent's inherent susceptibility to producing unreliable outputs. \textit{(ii) Inter-agent propagation modeling.} Hallucination diffusion along communication links is then modeled by considering inter-agent semantic influence and structural properties of the communication topology. \textit{(iii) System-level risk inference.} Finally, global risk aggregation is performed using a differentiable Noisy-OR inference over the agent communication graph, integrating intra- and inter-agent risk estimates to diagnose system-level hallucination risk. Extensive experiments on three datasets demonstrate that HalluProp effectively reduces MAS failure rates, with comparable performance to state-of-the-art post-hoc baselines. Specifically, it accurately localizes prospective failure risks with an average AUROC of $84.6\%$ and sub-second diagnosis latency ($65\times$ speedup over post-hoc methods), highlighting its practicality as an upstream risk screening in MAS.

The \textbf{key contributions} of this work are as follows:
\begin{itemize}[topsep=1.0pt, itemsep=-1.0pt]
    \item We investigate a new hallucination diagnosis strategy in MAS by shifting the focus from \textit{post-hoc} detection to \textit{pre-hoc} risk inference, enabling early localization and intervention of agent failure risks before MAS execution.
    \item We propose HalluProp, a propagation-aware hallucination diagnosis framework that aggregates intra- and inter-agent risks, enabling end-to-end systemic hallucination inference before response generation using a differentiable Noisy-OR inference over the communication graph.
    \item Extensive experiments show that HalluProp achieves an average AUROC of $84.6\%$ with sub-second latency, highlighting its practicality as an upstream risk-screening mechanism complementary to post-hoc methods.
\end{itemize}

%\vspace{-0.6em}
\section{Preliminary}
\label{preliminary}

\PP{Multi-Agent System}
A multi-agent system (MAS) can be modeled as a directed communication graph $\mathcal{M} = (\mathcal{V}, \mathcal{E})$, where $\mathcal{V} = \{v_1, \ldots, v_N\}$ denotes a set of LLM-based agents and $\mathcal{E} \in \{0,1\}^{N \times N}$ represents inter-agent communication links. Each agent $v_i$ is associated with a system prompt $p_i$ that specifies its role, capability, and behavioral constraints.
Given a user query $q$, agents collaboratively solve the task by exchanging intermediate responses along the links in $\mathcal{E}$. The connectivity encoded by $\mathcal{E}$ determines how information flows across agents and how errors may propagate in the system.

\PP{Hallucination in Multi-Agent System}
Hallucination refers to the generation of content that is factually incorrect, logically inconsistent, or unsupported by the given context. In multi-agent systems, hallucinations exhibit more complex characteristics. Beyond being produced independently by individual agents, hallucinations can propagate along the inter-agent communication chain. Errors generated by individual agents can spread and amplify through interactions, resulting in cascading system-level failures. Therefore, the risk of hallucinations is shaped not only by agent capabilities but also by agent roles and communication topology.

\PP{Pre-hoc \textit{vs.} Post-hoc}
Most existing hallucination detection methods adopt a \emph{post-hoc} strategy, diagnosing failures from observed agent responses and interactions after hallucinations have emerged, as illustrated in Figure~\ref{fig1}. In contrast, we advocate a \emph{pre-hoc} paradigm that estimates agent failure risks prior to inter-agent interaction. In MAS, pre-hoc analysis characterizes both intrinsic agent vulnerabilities and interaction-induced risks by (i) identifying hallucination tendencies within individual agents and (ii) estimating potential risk propagation along inter-agent communication. Instead of replacing runtime monitoring of long-horizon context drift, pre-hoc analysis provides an upstream safety signal for designing and refining risk-aware MAS architectures, complementing existing \emph{post-hoc} safeguards.

\PP{Problem Formulation}
We formulate the pre-hoc hallucination diagnosis in MAS as a latent risk diffusion problem over a directed communication graph. Let $H_i \in \{0,1\}$ indicate whether agent $v_i$ produces a hallucinated output under a given MAS configuration $(\{p_i\}_{i=1}^N, \mathcal{E})$. Our goal is to develop an approach that infers the hallucination risk of each agent before any response is generated:
\begin{align}
\small
h_i := \mathbb{P}\big(H_i = 1 \mid \{p_i\}_{i=1}^N, q, \mathcal{E}\big).  
\end{align}
Specifically, the hallucination risk in a MAS can be factorized into \emph{intra-agent intrinsic risk} and \emph{inter-agent propagation risk}. Intrinsic risk is represented as $\boldsymbol{\pi} = [\pi_1, \ldots, \pi_N]^\top \in [0,1]^N$, where $\pi_i$ captures the inherent tendency of agent $v_i$ to hallucinate, given its role specification and query. Propagation risk is modeled as a weighted adjacency matrix $\mathbf{T} \in [0,1]^{N \times N}$, where $T_{ij} = t_{j \rightarrow i}$ denotes the probability that hallucination propagates from agent $v_j$ to $v_i$ via communication links. Differentiable Noisy-OR inference is then performed over the communication graph based on $(\boldsymbol{\pi}, \mathbf{T})$ to derive the systemic hallucination risk vector $\mathbf{h} = [h_1, \ldots, h_N]^\top$, which provides a global diagnosis of hallucination risk by integrating node-level risks and edge-level propagation effects.

\begin{figure*}%[h]
    \centering
    \includegraphics[width=1.0\linewidth]{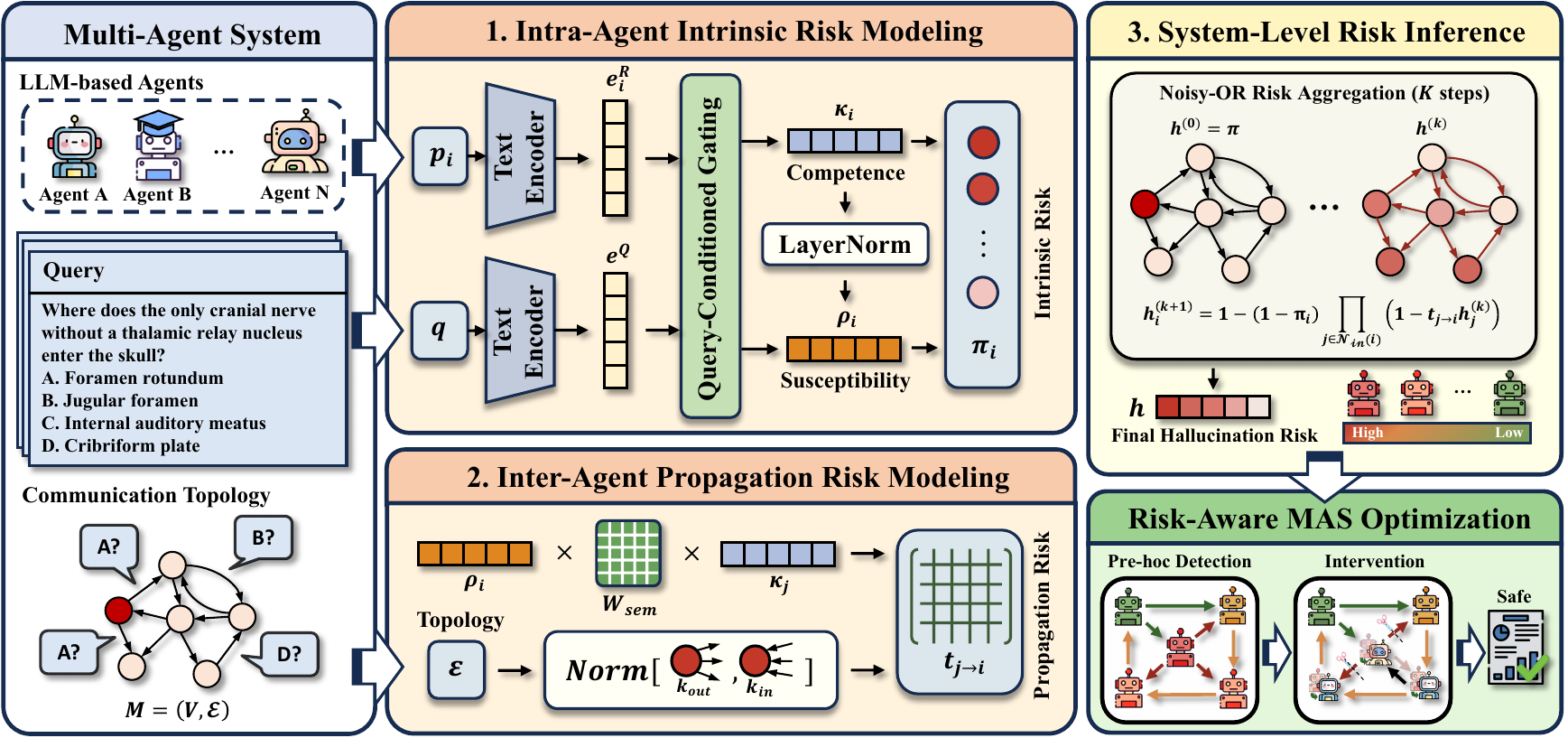}
    \vspace{-1.2em}
    \caption{The overall structure of HalluProp comprises three components: (1) \emph{intra-agent intrinsic risk modeling} that captures each agent's inherent hallucinations; (2) \emph{inter-agent propagation risk modeling} that identifies propagation risks among agent communications; and (3) \emph{system-level risk inference} that combines intra- and inter-agent risks to produce a final diagnosis.}
    \vspace{-0.4em}
    \label{fig:figure_overview}
\end{figure*}

\section{Method}
\label{method}

The overall architecture of the proposed framework is illustrated in Figure~\ref{fig:figure_overview}, which consists of three components: 
1) intra-agent intrinsic risk modeling, which characterizes the fine-grained semantic misalignment between role specifications and user queries to capture the inherent hallucination tendency of each agent before communication; 
2) inter-agent propagation risk modeling, which collectively considers inter-agent semantic influence and the structural properties of the communication topology to identify the hallucination propagation risk; 
and 3) system-level risk inference that aggregates the intra- and inter-agent risk via a differentiable Noisy-OR operator to produce a systemic hallucination diagnosis. In what follows, we elaborate on the three components.

\subsection{Intra-Agent Intrinsic Risk Modeling}
\label{sec:intrinsic}

Our first insight is to estimate the endogenous hallucination risk prior $\pi_i$ of each agent $v_i$ before inter-agent interaction occurs. This characterizes the probability of an agent producing unreliable content, which stems from the misalignment between its role specification \(p_i\) and user query \(q\). Such misalignment reflects if an agent's role-defined ability is compatible with the queried task and is a source of hallucinations.

\PP{Role-Query Semantic Encoding}
We begin by embedding the role specification \(p_i\) and the query \(q\) into a shared semantic space using a pretrained text encoder $\mathrm{Enc}(\cdot)$:
\begin{align}
\small
\mathbf{e}_i^{R} = \mathrm{Enc}(p_i), \quad
\mathbf{e}^{Q} = \mathrm{Enc}(q), \quad
\mathbf{e}_i^{R}, \mathbf{e}^{Q} \in \mathbb{R}^{d}.
\end{align}
Unlike task descriptions, role prompts often provide ambiguous  or misleading information. Therefore, global semantic similarity between $\mathbf{e}_i^{R}$ and $\mathbf{e}^{Q}$ is insufficient to capture the latent risk associated with each agent.

\PP{Query-Conditioned Gating}
To capture fine-grained correspondence between roles and queries, we introduce a \emph{query-conditioned gating mechanism} that adaptively filters out role-irrelevant features.
\begin{align}
\small
\mathbf{g}_i = \sigma\!\left(\mathbf{W}_{\mathrm{gate}}\big(\mathbf{e}_i^{R}\odot \mathbf{e}^{Q}\big) + \mathbf{b}_g\right),
\end{align}
where $\odot$ denotes element-wise multiplication that encodes dimension-wise semantic alignment. $\mathbf{W}_{\mathrm{gate}} \in \mathbb{R}^{d \times d}$ and $\mathbf{b}_g$ are learnable parameters, and $\sigma(\cdot)$ is the sigmoid function that maps the gating signal to $[0,1]$. The resulting gate $\mathbf{g}_i$ functions as a soft mask that emphasizes role dimensions relevant to the task while suppressing irrelevant components.

\PP{Competence-Susceptibility Decomposition}
Based upon the learned gate $\mathbf{g}_i$, the role representation is decomposed into two complementary subspaces:
\begin{align}
\small
\boldsymbol{\kappa}_i = \mathbf{e}_i^{R} \odot \mathbf{g}_i, \quad
\boldsymbol{\rho}_i = \mathrm{LayerNorm}\!\left(\mathbf{e}_i^{R} - \boldsymbol{\kappa}_i\right).
\end{align}
The \emph{competence vector} $\boldsymbol{\kappa}_i$ encodes role features that are aligned with the query and indicative of task-relevant capability. By contrast, the \emph{susceptibility vector} $\boldsymbol{\rho}_i$ aggregates residual role components that are weakly-related, conflicting, or potentially risk-inducing with respect to the query. This decomposition provides a structured representation of an agent's inherent propensity for hallucination. In practice, role prompts may be underspecified, overly generic, or adversarially crafted. Therefore, a holistic role-query similarity score may produce unstable estimates under prompt variation. Instead, the intrinsic risk module performs fine-grained dimension-wise relevance selection, allowing the model to retain task-aligned features while isolating residual parts that may induce hallucinations.
Finally, the intrinsic hallucination risk $\pi_i$ of each agent is modeled as follows:
\begin{align}
\small
\pi_i=\sigma\!\left(\mathbf{w}_{\rho}^{\top}\boldsymbol{\rho}_i-\mathbf{w}_{\kappa}^{\top}\boldsymbol{\kappa}_i+b_{\pi}\right),
\end{align}
where $\mathbf{w}_{\rho}$ and $\mathbf{w}_{\kappa}$ are learnable projection vectors that control the contributions of susceptibility and competence respectively, and $b_{\pi}$ is a bias term. Intuitively, such a scheme provides an interpretable estimate of an agent's intrinsic hallucination risk: an agent is more likely to hallucinate when its role exhibits strong risk-inducing residual traits (larger $\boldsymbol{\rho}_i$) while lacking sufficient query-aligned competence (smaller $\boldsymbol{\kappa}_i$).

\subsection{Inter-Agent Propagation Risk Modeling}
\label{sec:propagation}

Our second insight is that hallucinations in agent \(v_i\) may arise in MAS not only \emph{endogenously} from its own generation process, but also \emph{exogenously} through the adoption of erroneous content produced by other agents. 
Consequently, the intrinsic risk \(\pi_i\) of an agent alone is insufficient to diagnose hallucinations in MAS, where errors can propagate and amplify along inter-agent communication pathways.
To explicitly model this effect, we further estimate the \textit{inter-agent propagation risk} $t_{j \to i} \in [0,1]$ for each directed communication link $v_j \to v_i$, which measures the probability that unreliable information generated by $v_j$ successfully infects $v_i$.

\PP{Propagation Factor}
We believe that hallucination propagation in MAS is jointly governed by two complementary factors: 
\emph{(i) role semantics}, which determine the impact of sender and the susceptibility of receiver, and
\emph{(ii) communication topology}, which encodes the structural capacity of agents to broadcast or aggregate information. Accordingly, the propagation risk $t_{j \to i}$ is parameterized as follows:
\begin{align}
\small
\ell^{\mathrm{sem}}_{j\to i}
= \boldsymbol{\kappa}_j^{\top}\mathbf{W}_{\mathrm{sem}}\boldsymbol{\rho}_i,
\quad
t_{j\to i}=\sigma\!\Big(\gamma_{j,i}\cdot \ell^{\mathrm{sem}}_{j\to i}+b_t\Big).
\end{align}
For role semantics, we reuse \(\boldsymbol{\kappa}_j\) and \(\boldsymbol{\rho}_i\). 
\(\boldsymbol{\kappa}_j\) represents the sender's task-aligned capability to provide reliable information, while $\boldsymbol{\rho}_i$ characterizes the receiver's susceptibility to unreliable or misaligned information. The learnable matrix $\mathbf{W}_{\mathrm{sem}}$ captures their semantic compatibility and interaction effects, with \(b_t\) as a learnable bias term.

\PP{Topology Effect}
To incorporate the communication topology with minimal overhead, we extract lightweight structural features based on node degrees:
\begin{align}
\small
\gamma_{j,i} = \phi\big(\hat{k}_{\mathrm{out}}(j), \hat{k}_{\mathrm{in}}(i)\big),
\end{align}
where $\hat{k}_{\mathrm{out}}(j)$ and $\hat{k}_{\mathrm{in}}(i)$ denote the normalized out-degree and in-degree of sender $v_j$ and receiver $v_i$ respectively, and $\phi(\cdot)$ is parameterized by an MLP whose output is mapped to $[1, \infty)$. It is worth mentioning that propagation risk increases when a highly influential sender interacts with a susceptible receiver, and is amplified ($\gamma_{j,i} \ge 1$) by strong broadcasting or aggregation capacity reflected in the topology. To ensure structural validity, propagation is restricted to existing communication links by enforcing $t_{j \to i} = 0$ if $(v_j \to v_i) \notin \mathcal{E}$.

\subsection{System-Level Risk Inference}
\label{system-level}

Given the intra-agent intrinsic risks $\boldsymbol{\pi}$ and inter-agent propagation risks $\mathbf{T}$, HalluProp performs \emph{system-level cascading hallucination inference} by modeling agent failure as a multi-source triggering process over the MAS topology structure. The goal is to deduce \emph{final hallucination risk} for all agents, denoted by $\mathbf{h} = [h_1, \ldots, h_N]^\top$, which integrates both endogenous generation risk and exogenous propagation effects.

\PP{Noisy-OR Risk Aggregation}
An agent $v_i$ remains non-hallucinating if and only if it neither hallucinates intrinsically nor is infected by any upstream agent through communication links. Under the standard conditional-independence assumption adopted by the differentiable Noisy-OR mechanism, the final hallucination risk of agent $v_i$ is given by
\begin{align}
\small
h_i = 1-(1-\pi_i)\prod_{j \in \mathcal{N}_{\mathrm{in}}(i)}\bigl(1 - t_{j \to i} \, h_j \bigr),
\end{align}
where $\mathcal{N}_{\mathrm{in}}(i)$ denotes the set of in-neighbors of $v_i$ and $h_j$ is the final hallucination risk of $v_j$. Each factor $(1 - t_{j \to i} h_j)$ represents the probability that agent $v_j$ fails to transmit hallucinated content to $v_i$, while the product aggregates independent failure events. Taking the complement gives the probability that hallucination arises from at least one source, either intrinsically or via propagation.

\PP{Fixed-Point Inference} 
Since a MAS communication graph typically contains cycles, closed-form solutions are generally intractable. We therefore compute $\mathbf{h}$ via $K$ steps of fixed-point iteration, where at each step $k \in \{0, \dots, K-1\}$:
\begin{align}
\small
h_i^{(k+1)} = 1-(1-\pi_i)\prod_{j\in\mathcal{N}_{\mathrm{in}}(i)}
\bigl(1-t_{j\to i}h_j^{(k)}\bigr).
\end{align}
Initialization with $h_i^{(0)} = \pi_i$ corresponds to the hallucination risk in the absence of inter-agent propagation, while successive iterations gradually incorporate higher-order diffusion effects.
The fixed-point update is fully differentiable with respect to both $\boldsymbol{\pi}$ and $\mathbf{T}$, enabling end-to-end training of the entire framework. After $K$ iterations, the final risk $\mathbf{h} = \mathbf{h}^{(K)}$ illustrates a global diagnosis of hallucination risk over the entire system. Notably, we have put the theoretical analysis of the proposed Noisy-OR mechanism in the appendix.

\subsection{Two-stage Training Process}

We adopt a two-stage training paradigm that reflects the hierarchical risk formation process in MAS, where hallucination risks may arise from individual agents and subsequently emerge via inter-agent interactions. Specifically, the first stage learns agent-level intrinsic risk priors, while the second stage optimizes interaction-induced risk under communication graph-level supervision. 

In the \textit{first} stage, we train the intrinsic risk module while keeping the text encoder frozen. Each agent's intrinsic hallucination risk $\pi_i$ is supervised by binary cross-entropy:
\begin{align}
\small
\mathcal{L}_{\mathrm{s1}} = \mathrm{BCE}\!\left(\boldsymbol{\pi}, \mathbf{y}^{\mathrm{node}}\right),
\end{align}
where $\mathbf{y}^{\mathrm{node}}$ denotes node-level hallucination labels before communication, which characterize each agent's intrinsic tendency to generate unreliable information.
In the \textit{second} stage, we initialize the model from the checkpoint of first stage to optimize the propagation risk module using graph-level supervision. The system-level hallucination risks $\mathbf{h}^{(K)}$ are obtained via $K$ iterations of differentiable Noisy-OR fixed-point propagation. The optimization objective is defined as:
\begin{align}
\small
\mathcal{L}_{\mathrm{s2}} = \mathrm{BCE}\big(\mathbf{h}^{(K)}, \mathbf{y}^{\mathrm{graph}}\big) + \lambda_{\mathrm{anc}}\,\mathcal{L}_{\mathrm{s1}},
\end{align}
where $\mathbf{y}^{\mathrm{graph}}$ represents post-communication hallucination labels reflecting interaction-induced risk. The anchor term $\lambda_{\mathrm{anc}}$ preserves the intrinsic risk estimation during graph-level optimization, enabling the propagation matrix $\mathbf{T}$ to capture additional risk introduced by agent interactions. The entire framework is differentiable and supports communication graphs with varying numbers of agents using node masking.

\section{Experiment}

In this section, we conduct extensive experiments to evaluate HalluProp, seeking to answer the following questions:
\begin{itemize}[topsep=2.0pt, itemsep=0.5pt, left=0.0em]
    \item \textbf{RQ1:} Can HalluProp accurately identify high-risk agents in MAS? Can it effectively guide early intervention?
    \item \textbf{RQ2:} How efficient and robust is HalluProp across varying MAS scales and communication topologies?
    \item \textbf{RQ3:} How do individual components and design choices contribute to hallucination risk inference?
\end{itemize}

\subsection{Experimental Setup}

\PP{Datasets}
We evaluate HalluProp on three public benchmarks that span general knowledge, mathematical reasoning, and domain-specific expertise: (1) \textit{MMLU}~\cite{hendrycks2021measuring}, covering 57 subjects across a wide range of domains; (2) \textit{MATH}~\cite{hendrycks2021measuring}, consisting of challenging mathematical problems; and (3) \textit{MedQA}~\cite{jin2021disease}, a specialized biomedical question-answering benchmark.

\PP{Baselines} 
We compare the proposed framework against four representative baselines. (1) \emph{LLM Debate}~\cite{du2023improving}, which employs multi-agent debate to correct errors; (2) \emph{Inspector}~\cite{huang2024resilience}, which introduces a dedicated verifier agent to review and correct responses; (3) \emph{SelfCheckGPT}~\cite{manakul2023selfcheckgpt}, which detects hallucinations by measuring inconsistencies across multiple sampled outputs; and (4) \emph{GUARDIAN}~\cite{zhou2026guardian}, which models MAS as a temporal communication graph to capture dynamic error patterns. Notably, all baselines follow a post-hoc detection paradigm and require the completion of full MAS execution.

\PP{Metrics} 
To evaluate the performance of the proposed framework, we adopt the following metrics:
\begin{itemize}[topsep=0.0em, itemsep=0.0pt, left=0.0em]
    \item \textbf{Risk Inference and Intervention Utility (RQ1).} 
    We measure the alignment between predicted and ground-truth hallucination labels using AUROC. To assess localization accuracy and coverage, we report Hit@$k$, Precision@$k$, and Recall@$k$ over the top-$k$ ranked agents, with $k \in \{1, 3\}$. To show practical utility, we report post-intervention accuracy.
    \item \textbf{Efficiency, Scalability, and Structural Generalization (RQ2).} 
    We evaluate computational efficiency by measuring total diagnosis time per query and the speedup ratio relative to full MAS execution. Scalability is assessed by tracking AUROC stability under different numbers of agents. To examine structural generalization, we evaluate HalluProp under both random topologies and canonical engineered workflows, including star, chain, and hierarchical.
\end{itemize}

\PP{MAS Configuration} 
We set up multi-agent systems using four LLM backbones: the closed-source GPT-3.5-turbo and GPT-4o~\cite{achiam2023gpt}, and the open-source Llama-3-8B-Instruct~\cite{grattafiori2024llama} and Qwen-2-7B-Instruct~\cite{team2024qwen2}. To balance stability and collaborative diversity, we set the decoding temperature to $0.7$ across all models. For each dataset instance, we construct an interaction graph with a randomized topology, where agents are assigned distinct roles via role-specific system prompts. To ensure comprehensive evaluation, we instantiate agents using diverse role descriptions, including clean and noisy roles with irrelevant or misleading content. This reflects practical MAS settings where agent roles can be underspecified, vague, or adversarially phrased. Agent interactions proceed for three rounds to enable sufficient information exchange. To ensure structural diversity, we uniformly sample inter-agent connection probabilities from $25\%$, $50\%$, $75\%$, and $100\%$, yielding a broad spectrum of MAS topologies. Finally, the ground-truth hallucination label $y_i \in \{0, 1\}$ is determined by comparing each agent's output against the reference answer.

\renewcommand{\arraystretch}{0.65}
\begin{table}[t]
\centering
\fontsize{6.2pt}{6.6pt}\selectfont
\setlength{\tabcolsep}{0.5pt}

\begin{tabular*}{0.98\columnwidth}{
@{\extracolsep{\fill}}llcccccc@{}
}
\toprule

\multirow{2}{*}{\textbf{Model}}
& \multirow{2}{*}{\textbf{Dataset}}
& \multicolumn{6}{c}{\textbf{Metric}} \\

\cmidrule(l{0.25em}r{0.65em}){3-8}

& &
\textbf{AUROC}
& \textbf{Hit@1}
& \textbf{Rec@1}
& \textbf{Hit@3}
& \textbf{Prec@3}
& \textbf{Rec@3} \\

\midrule

\multirow{3}{*}{GPT-3.5-turbo}
& MMLU
& 89.1 & 86.4 & 44.2 & 92.7 & 79.6 & 80.1 \\
& MATH
& 88.5 & 85.1 & 43.3 & 91.1 & 78.0 & 79.8 \\
& MedQA
& 76.8 & 80.4 & 38.1 & 91.7 & 76.9 & 70.2 \\

\cmidrule(l{0.08em}r{0.08em}){1-8}

\multirow{3}{*}{GPT-4o}
& MMLU
& 91.5 & 89.2 & 48.8 & 94.5 & 82.1 & 84.5 \\
& MATH
& 91.0 & 88.5 & 47.5 & 93.8 & 81.0 & 83.2 \\
& MedQA
& 81.2 & 84.1 & 42.6 & 93.5 & 79.8 & 75.4 \\

\cmidrule(l{0.08em}r{0.08em}){1-8}

\multirow{3}{*}{Llama-3-8B}
& MMLU
& 86.8 & 84.2 & 40.8 & 90.8 & 77.1 & 76.5 \\
& MATH
& 86.2 & 82.8 & 39.5 & 89.5 & 75.8 & 75.8 \\
& MedQA
& 74.5 & 78.5 & 34.2 & 89.2 & 74.5 & 66.5 \\

\cmidrule(l{0.08em}r{0.08em}){1-8}

\multirow{3}{*}{Qwen-2-7B}
& MMLU
& 87.5 & 85.1 & 41.5 & 91.2 & 78.2 & 77.5 \\
& MATH
& 87.2 & 84.0 & 41.0 & 90.2 & 77.0 & 77.2 \\
& MedQA
& 75.1 & 79.2 & 35.1 & 89.8 & 75.5 & 67.8 \\

\midrule

\multicolumn{2}{c}{\textbf{Average}}
& \textbf{84.6}
& \textbf{83.9}
& \textbf{41.4}
& \textbf{91.5}
& \textbf{77.9}
& \textbf{76.2} \\

\bottomrule
\end{tabular*}

\vspace{-0.6em}
\caption{Hallucination detection performance (\%) evaluated on three datasets. Our method demonstrates robust identification and precise localization across LLMs. Prec and Rec denote precision and recall, respectively.}
\label{tab:rq1_main_results}
\vspace{-0.8em}
\end{table}

\begin{table*}[t]
\centering
\renewcommand{\arraystretch}{0.82}
\setlength{\tabcolsep}{1.5pt}

{\fontsize{8.0pt}{7.8pt}\selectfont
\begin{tabular*}{\textwidth}{
@{\extracolsep{\fill}}l*{13}{c}@{}
}
    \toprule
    \multirow{2}{*}{\textbf{Method}}
    & \multicolumn{3}{c}{\textbf{GPT-3.5-turbo}}
    & \multicolumn{3}{c}{\textbf{GPT-4o}}
    & \multicolumn{3}{c}{\textbf{Llama-3-8B}}
    & \multicolumn{3}{c}{\textbf{Qwen-2-7B}}
    & \multirow{2}{*}{\textbf{Average}} \\

    \cmidrule(lr){2-4}
    \cmidrule(lr){5-7}
    \cmidrule(lr){8-10}
    \cmidrule(lr){11-13}

    & \textbf{MMLU} & \textbf{MATH} & \textbf{MedQA}
    & \textbf{MMLU} & \textbf{MATH} & \textbf{MedQA}
    & \textbf{MMLU} & \textbf{MATH} & \textbf{MedQA}
    & \textbf{MMLU} & \textbf{MATH} & \textbf{MedQA}
    & \\

    \midrule

    LLM Debate
    & 55.2 & 48.1 & 44.1
    & 76.6 & 70.9 & 71.3
    & 58.4 & 51.9 & 37.5
    & 60.1 & 54.6 & 40.1
    & 55.7 \\

    Inspector
    & 51.9 & 49.5 & 39.5
    & 73.1 & 72.4 & 69.1
    & 57.2 & 53.1 & 35.2
    & 58.4 & 56.2 & 37.4
    & 54.4 \\

    SelfCheckGPT
    & 53.4 & 50.2 & 46.2
    & 77.9 & 73.8 & 69.8
    & 57.1 & 54.5 & 39.1
    & 59.5 & 57.8 & 41.0
    & 56.7 \\

    GUARDIAN
    & 66.8 & 58.5 & 53.6
    & 83.5 & 78.5 & 75.2
    & 64.2 & 59.2 & 45.4
    & 66.8 & 62.1 & 47.8
    & 63.5 \\

    \midrule

    HalluProp-Pruning (\textbf{ours})
    & 64.4 & 60.8 & 53.3
    & 80.9 & 77.2 & 76.5
    & 65.8 & 58.8 & 46.8
    & 67.5 & 62.5 & 49.2
    & 63.6 \\

    HalluProp-Refinement (\textbf{ours})
    & \textbf{70.2} & \textbf{62.2} & \textbf{59.8}
    & \textbf{85.7} & \textbf{80.4} & \textbf{79.4}
    & \textbf{68.5} & \textbf{64.5} & \textbf{51.2}
    & \textbf{70.4} & \textbf{64.8} & \textbf{54.5}
    & \textbf{67.6} \\

    \bottomrule
\end{tabular*}
}

\vspace{-0.5em}
\caption{Effectiveness of hallucination mitigation strategies guided by HalluProp. We report the post-intervention accuracy (\%) compared to four representative baselines across different LLMs and datasets.}
\label{tab:mitigation_results}
\vspace{-0.4em}
\end{table*}

\subsection{RQ1: Risk Inference and Intervention Capability}

We evaluate the alignment between the predicted hallucination risk scores $h_i$ and the ground-truth labels $y_i$ across multiple LLMs and datasets. To confirm the practical effectiveness of HalluProp in alleviating hallucinations, we apply intervention strategies guided by the predicted risk scores $h_i$.
\begin{itemize}[topsep=1.0pt, itemsep=-0.5pt, left=0.0em]
    \item \textbf{Pruning:} disabling the highest-risk agent by removing its outgoing communication edges.
    \item \textbf{Refinement:} reassigning the highest-risk agent with a more cautious role prompt.
\end{itemize}

\PP{Precise Identification and Localization} 
Our framework demonstrates strong capability in both risk identification and localization. As shown in Table~\ref{tab:rq1_main_results}, it achieves an average AUROC of $84.6\%$, indicating a reliable discrimination between hallucinating and faithful agent behaviors. More importantly, HalluProp excels at error localization. In most configurations, Hit@1 exceeds $80\%$, while Recall@3 reaches $76.2\%$, providing a balance between localization precision and coverage. Despite differences in architecture, parameter scale, and knowledge domain, it maintains consistent performance across both closed-source models (GPT series) and open-source models (Llama-3 and Qwen-2) in various datasets. Overall, these results demonstrate effective pre-execution risk identification in multi-agent systems.

\begin{figure}[t]
    \centering
    \begin{subfigure}[t]{1\linewidth}
        \centering
        \includegraphics[width=\linewidth]{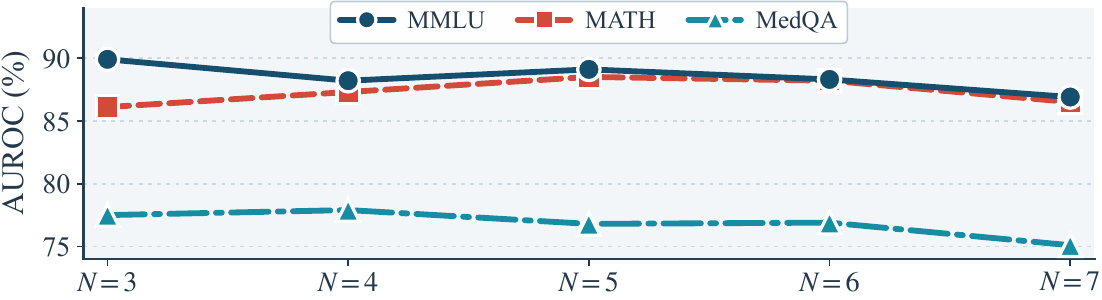}
        \vspace{-1.0em}
        \caption{Scalability across different numbers of agents}
        \label{fig:scalability_sub}
    \end{subfigure}
    \hfill
    \begin{subfigure}[t]{1\linewidth}
        \centering
        \includegraphics[width=\linewidth]{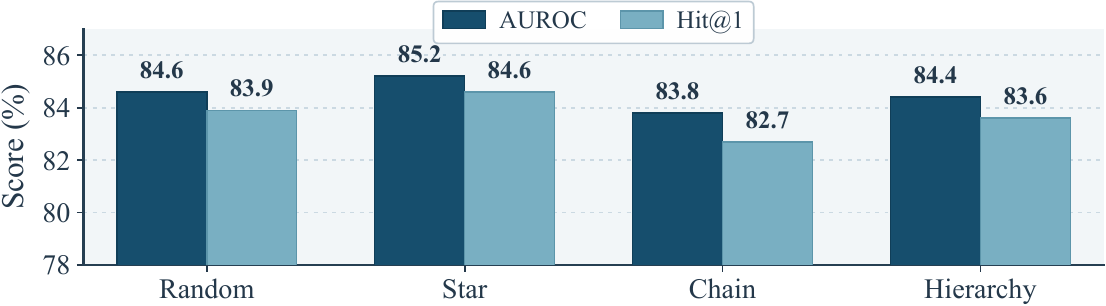}
        \vspace{-1.0em}
        \caption{Robustness across different topologies}
        \label{fig:topology_sub}
    \end{subfigure}
    \vspace{-0.2em}
    \caption{Robustness of HalluProp with different MAS configurations. (a) Scalability across varying numbers of agents. (b) Performance under different structured topologies.}
    \label{fig:rq2_robustness}
    \vspace{-0.4em}
\end{figure}

\PP{Improvements in System Reliability} 
We evaluate system performance after applying interventions, with results summarized in Table~\ref{tab:mitigation_results}. Overall, risk-aware interventions guided by HalluProp lead to reliability gains comparable to strong post-hoc mitigation baselines across different models and datasets. In particular, the pruning strategy achieves an average accuracy of $63.6\%$, closely matching the state-of-the-art baselines, GUARDIAN, at $63.5\%$. Notably, this performance is achieved using risks inferred before inter-agent interaction, without observing agent responses or interaction traces, highlighting the potential of pre-hoc risk inference for building more reliable MAS as an upstream screening mechanism.

\PP{Benefits of Refinement Strategy}
We further observe that refinement strategy consistently outperforms pruning across settings, improving the average accuracy from $63.6\%$ to $67.6\%$. For example, with the Llama-3-8B/MATH configuration, pruning attains $58.8\%$, slightly below the strongest baseline GUARDIAN at $59.2\%$, whereas refinement achieves a higher accuracy of $64.5\%$. These results suggest that directly removing an agent node may inadvertently disrupt useful information flow or reduce collaborative diversity in multi-agent systems. In contrast, refinement preserves the agent’s contribution while encouraging more cautious and reliable reasoning. These findings demonstrate that pre-hoc risk estimates can support not only structural intervention but also more effective behavioral regulation in MAS.

\subsection{RQ2: Efficiency, Scalability, and Generalization}

% To validate the feasibility of deploying HalluProp in real-world settings,  we evaluate it based on three aspects: computational efficiency, scalability, and structural generalization.

\PP{Real-time Inference Efficiency}
We benchmark the inference latency of HalluProp against four representative baseline methods using GPT-3.5-turbo. Since all baselines are \textit{post-hoc}, they inevitably incur the full cost of MAS execution, including multi-round LLM generation and communication overhead. In contrast, HalluProp operates exclusively on the static communication graph, thus avoiding the costly sequential generation of tokens and network I/O during agent interactions. As shown in Table~\ref{tab:efficiency}, HalluProp completes hallucination diagnosis in less than one second, yielding over a $65\times$ speedup relative to post-hoc methods. This efficiency enables HalluProp to serve as a lightweight firewall, facilitating real-time hallucination detection before execution.

\PP{Scalability Across MAS Scales} 
To investigate scalability under varying MAS scales, we evaluate HalluProp with different numbers of agents ($N \in \{3,4,5,6,7\}$). As illustrated in Figure~\ref{fig:rq2_robustness}, HalluProp exhibits stable performance despite increasing collaboration size. Across all benchmarks, AUROC remains within narrow ranges ($85\%$-$90\%$ on MMLU and MATH, $75\%$-$80\%$ on MedQA), indicating minimal sensitivity to scale. These results suggest that HalluProp captures scale-invariant patterns of hallucination propagation, ensuring robust performance across dynamic MAS configurations.

\PP{Performance Across Structured Workflows}
Unlike random communication topologies, real-world MAS often adopt engineered workflows such as star, chain, and hierarchical. Therefore, we evaluate HalluProp across these canonical topologies. As shown in Figure~\ref{fig:rq2_robustness}, HalluProp maintains stable performance, with both AUROC and Hit@1 differing by less than $2\%$ from the random topology setting. This indicates that HalluProp captures generalizable structural patterns, instead of simply overfitting to specific topology structures.

\begin{table}[t]
\centering
\renewcommand{\arraystretch}{0.78}
\setlength{\tabcolsep}{1.8pt}
{\fontsize{8.5pt}{8.3pt}\selectfont

\begin{tabular*}{\columnwidth}{
@{\extracolsep{\fill}}llcc@{}
}
\toprule
\textbf{Method}
& \textbf{Type}
& \textbf{Total Time Cost}
& \textbf{Speedup} \\
\midrule

LLM Debate
& \textit{Post-hoc}
& 65.5s
& 1.0$\times$ \\

Inspector
& \textit{Post-hoc}
& 158.1s
& 0.4$\times$ \\

SelfCheckGPT
& \textit{Post-hoc}
& 308.4s
& 0.2$\times$ \\

GUARDIAN
& \textit{Post-hoc}
& 53.2s
& 1.2$\times$ \\

\midrule

HalluProp (\textbf{ours})
& \textit{Pre-hoc}
& \textbf{$<$ 1s} 
& \textbf{$>$ 65$\times$} \\

\bottomrule
\end{tabular*}
}

\vspace{-0.4em}
\caption{Efficiency comparison on GPT-3.5-turbo.}
\label{tab:efficiency}
\vspace{-0.6em}
\end{table}

\subsection{RQ3: Ablation Study and Configuration Analysis}
\label{sec:ablation}

\PP{Intra-Agent Intrinsic Risk Modeling}
To examine whether the proposed intrinsic-risk module provides a stronger signal, we replace the intrinsic risk $\pi_i$ with a similarity-based estimate computed directly from the frozen encoder representations, which uses the holistic role-query alignment as the intrinsic risk prior. The observed performance drop (approximately $25\%$) in Table~\ref{tab:ablation} indicates that coarse similarity is insufficient to capture the tendency of agents to hallucinate. This confirms that our query-conditioned competence-susceptibility modeling extracts fine-grained risk-inducing features, leading to more informative pre-execution risk estimation.

\PP{Topology Features}
To measure the impact of communication structure, we exclude the topology factor $\gamma_{j,i}$ from the propagation formulation, leaving only the semantic interaction term $\ell^{\mathrm{sem}}_{j\to i}$. As shown in Table~\ref{tab:ablation}, this leads to a significant performance decrease, with AUROC decreasing by approximately $15\%$. This result highlights that the structural properties of the interaction graph (e.g., node degree and connectivity patterns) amplify risk diffusion, and removing topological information weakens the identification of high-impact broadcasters, reducing the system-level hallucination diagnosis accuracy.

\begin{table}[t]
\centering
\renewcommand{\arraystretch}{0.82}
\setlength{\tabcolsep}{2.5pt}

{\fontsize{8.3pt}{8.3pt}\selectfont

\begin{tabularx}{0.99\columnwidth}{
@{}
X
>{\centering\arraybackslash}p{0.35\columnwidth}
@{}
}
\toprule

\textbf{Model Variant}
& \textbf{Average AUROC} \\

\midrule

\textbf{HalluProp}
& \textbf{84.8} \\

\cmidrule{1-2}

\textit{w/o Intra-Agent Risk Modeling}
& \makebox[\linewidth][c]{%
    \makebox[2.4em][r]{61.2}%
    \hspace{0.45em}%
    \makebox[4.8em][l]{%
        {\scriptsize\bfseries\itshape
        \(\left(\downarrow\!27.8\%\right)\)}
    }%
} \\

\textit{w/o Topology Features}
& \makebox[\linewidth][c]{%
    \makebox[2.4em][r]{72.1}%
    \hspace{0.45em}%
    \makebox[4.8em][l]{%
        {\scriptsize\bfseries\itshape
        \(\left(\downarrow\!15.0\%\right)\)}
    }%
} \\

\textit{w/o Inter-Agent Propagation}
& \makebox[\linewidth][c]{%
    \makebox[2.4em][r]{66.9}%
    \hspace{0.45em}%
    \makebox[4.8em][l]{%
        {\scriptsize\bfseries\itshape
        \(\left(\downarrow\!21.1\%\right)\)}
    }%
} \\

\bottomrule
\end{tabularx}
}

\vspace{-0.3em}
\caption{Ablation study of HalluProp on GPT-3.5-turbo (\%).}
\label{tab:ablation}
\vspace{-0.2em}
\end{table}

% \definecolor{DropRed}{HTML}{C44237}

% \begin{table}[t]
% \centering
% \renewcommand{\arraystretch}{0.82}
% \setlength{\tabcolsep}{2.5pt}

% {\fontsize{8.3pt}{8.3pt}\selectfont

% \begin{tabularx}{0.99\columnwidth}{
% @{}
% X
% >{\centering\arraybackslash}p{0.35\columnwidth}
% @{}
% }
% \toprule

% \textbf{Model Variant}
% & \textbf{Average AUROC} \\

% \midrule

% \textbf{HalluProp}
% & \textbf{84.8} \\

% \cmidrule{1-2}

% \textit{w/o Intra-Agent Risk Modeling}
% & \makebox[\linewidth][c]{%
%     \makebox[2.4em][r]{61.2}%
%     \hspace{0.45em}%
%     \makebox[4.2em][l]{%
%         {\scriptsize\textcolor{DropRed}{\(\downarrow\!27.8\%\)}}
%     }%
% } \\

% \textit{w/o Topology Features}
% & \makebox[\linewidth][c]{%
%     \makebox[2.4em][r]{72.1}%
%     \hspace{0.45em}%
%     \makebox[4.2em][l]{%
%         {\scriptsize\textcolor{DropRed}{\(\downarrow\!15.0\%\)}}
%     }%
% } \\

% \textit{w/o Inter-Agent Propagation}
% & \makebox[\linewidth][c]{%
%     \makebox[2.4em][r]{66.9}%
%     \hspace{0.45em}%
%     \makebox[4.2em][l]{%
%         {\scriptsize\textcolor{DropRed}{\(\downarrow\!21.1\%\)}}
%     }%
% } \\

% \bottomrule
% \end{tabularx}
% }

% \vspace{-0.3em}
% \caption{Ablation study of HalluProp on GPT-3.5-turbo (\%).}
% \label{tab:ablation}
% \vspace{-0.2em}
% \end{table}

\PP{Inter-Agent Propagation Module}
To evaluate the necessity of modeling inter-agent influence, we disable the propagation module by enforcing $t_{j \to i}=0$ for all communication links. In this setting, the system-level risk $h_i$ collapses to the intra-agent intrinsic risk $\pi_i$. As shown in Table~\ref{tab:ablation}, this results in a substantial degradation in performance, with AUROC dropping by approximately $20\%$. These findings provide strong empirical evidence that hallucinations in MAS are not merely isolated intrinsic errors, but are significantly amplified through inter-agent propagation. Consequently, relying solely on intra-agent risk estimation is insufficient to capture cascading hallucination dynamics at the system level.

\begin{figure}[t]
\centering

\includegraphics[
    width=0.99\columnwidth,
    trim=2 2 2 2,
    clip
]{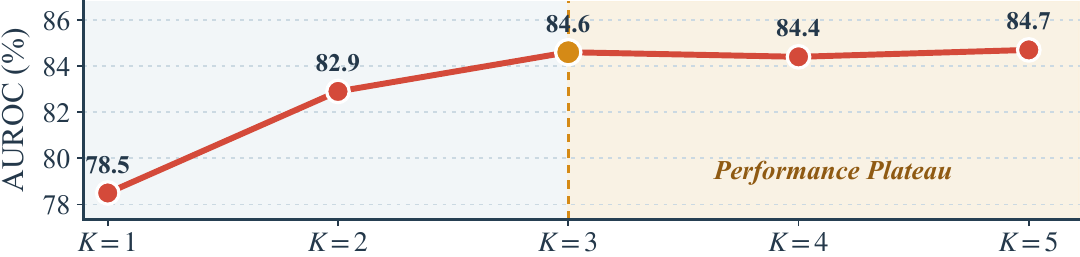}

\vspace{-0.1em}
\caption{Sensitivity to iteration depth $K$.}
\label{fig:k_sensitivity}
\vspace{-0.6em}
\end{figure}

\PP{Impact of Fixed-Point Inference Depth}
We further analyze the effect of the fixed-point iteration depth $K$, which controls how many propagation steps are incorporated into system-level inference. As shown in Figure~\ref{fig:k_sensitivity}, one step inference underestimates multi-hop hallucination diffusion, achieving an average AUROC of $78.5$. Increasing $K$ to $2$ improves AUROC to $82.9$, and $K=3$ further reaches $84.6$. After that, the performance saturates, with less than $0.3$ variation for $K \in \{3,4,5\}$. These results indicate that HalluProp benefits from multi-step propagation while remaining robust to the choice of $K$ once sufficient propagation depth is reached.

\PP{Sensitivity to Text Encoders}
Finally, we examine whether HalluProp relies on a specific text encoder. We replace the default encoder with several off-the-shelf sentence encoders, while keeping downstream modules unchanged. As shown in Table~\ref{tab:encoder_sensitivity}, HalluProp exhibits consistent performance across different encoders. These results suggest that the effectiveness of HalluProp mainly comes from downstream components, rather than specific representation backbone.

\section{Related Work}

\PP{Hallucination Detection in LLMs}
With the rapid adoption of large language models (LLMs), hallucination has emerged as a critical threat to reliability and safety. 
Existing detection approaches for LLMs can be broadly categorized into four types.
Retrieval-based methods align model outputs with retrieved information from a reliable knowledge base~\cite{min2023factscore,semnani2023wikichat,tang2024minicheck,zhang2025knowhalu}. 
Reasoning-based methods prompt models to self-examine generated responses via structured reasoning or counterfactual evidence~\cite{chen2023felm,dhuliawala2023chain,wei2022chain}. 
Uncertainty- and representation-based methods estimate hallucination risk using token-level uncertainty~\cite{yoffe2025debunc,hou2025probabilistic,varshney2023stitch,fadeeva2024fact} or internal hidden representations~\cite{azaria2023internal,chen2023inside,su2024unsupervised,sriramanan2024llm}. 
Consistency-based methods detect hallucinations by measuring inconsistencies across multiple sampled outputs for the same input~\cite{manakul2023selfcheckgpt,cao2023autohall,mundler2023self}. 
Notably, most existing methods identify hallucinations only after LLMs' response generation using external evidence or intrinsic signals.

\begin{table}[t]
\centering
\renewcommand{\arraystretch}{0.80}
\setlength{\tabcolsep}{1.8pt}
{\fontsize{8.5pt}{8.5pt}\selectfont

\begin{tabular*}{0.99\columnwidth}{
@{\extracolsep{\fill}}lccc@{}
}
\toprule
\textbf{Text Encoder}
& \textbf{AUROC}
& \textbf{Hit@1}
& \textbf{Rec@3} \\
\midrule

bert-base-uncased \textbf{(default)}
& 84.6
& 83.9
& 76.2 \\

e5-base-v2
& 83.8
& 82.1
& 75.3 \\

bge-base-en-v1.5
& 85.1
& 85.3
& 77.8 \\

all-MiniLM-L6-v2
& 80.7
& 78.9
& 74.1 \\

\bottomrule
\end{tabular*}
}

\vspace{-0.6em}
\caption{Sensitivity analysis with different text encoders (\%).}
\label{tab:encoder_sensitivity}
\vspace{-0.9em}
\end{table}

\PP{Hallucination Identification in MAS}
Multi-agent systems are increasingly scaling single-agent capabilities, yet the complex interactions introduce new hallucination risks~\cite{guo2024large}. In particular, hallucinations may propagate and amplify through inter-agent communication, resulting in cascades and collective failures~\cite{yoffe2025debunc,zhang2024language}. To identify and mitigate such risks, existing studies fall into two categories. Collaborative correction methods rely on multi-round debates, cross-checking, or majority voting to rectify errors after they occur~\cite{du2023improving,huang2024resilience,feng2024don,mehta2024improving}. Structure-aware approaches model MAS as a communication graph and analyze how topology shapes error propagation and amplification~\cite{shen2025understanding,zhou2026guardian}. However, these methods largely follow a post-hoc paradigm, detecting hallucinations only after responses are generated. In contrast, we take a pre-hoc method to infer hallucination that models both intra-agent intrinsic risk and inter-agent propagation risk without executing task interactions, providing interpretable guidance for role assignment and communication-structure design.

\section{Conclusion}

In this work, we investigate a new paradigm for hallucination detection in multi-agent systems by diagnosing potential hallucination risks prior to MAS execution. Particularly, we present HalluProp, a propagation-aware hallucination inference framework, which collaboratively combines intra-agent intrinsic risk and inter-agent propagation risks, enabling end-to-end differentiable system-level diagnosis via Noisy-OR inference. Empirical evaluations demonstrate that HalluProp achieves comparable post-intervention accuracy to state-of-the-art baselines. We believe our work provides new insights into hallucination risk modeling in multi-agent systems and opens new directions for proactive hallucination mitigation.

% \bibliography{aaai2027}

% \newpage

\newpage
\appendix

\twocolumn

\section{Appendix}

\subsection{Theoretical Analysis}
\label{sec:theoretical_analysis}

In this section, we provide a rigorous justification for the proposed Noisy-OR mechanism in HalluProp. Specifically, we analyze this differentiable formulation from three perspectives: its probabilistic grounding in causal independence, the convergence guarantees of the inference process, and the favorable gradient properties for optimization.

\PP{Causal Independence}
The Noisy-OR mechanism is theoretically based onthe assumption of \textit{causal independence} among failure factors. An agent $v_i$ remains reliable (i.e., non-hallucinating) if and only if it is intrinsically safe, and successfully immune to errors from all neighbors. Assuming that these safety conditions are independent, the final hallucination risk $h_i$ of $v_i$ is derived as the complement of the joint reliability probability:
\begin{equation}
\label{eq:noisy_or_logic}
h_i = 1 - \underbrace{(1-\pi_i)}_{\text{Intrinsic Safety}} \prod_{j \in \mathcal{N}_{\text{in}}(i)} \underbrace{(1 - t_{j \to i} h_j)}_{\text{Propagation Safety}}.
\end{equation}
This captures the ``any-cause'' logic of error propagation, whereby a single failure in the dependency chain compromises the node, while maintaining a smooth, differentiable manifold for learning, allowing the model to implicitly handle uncertainty and noise in multi-agent communications.

\PP{Convergence Guarantee}
Since the communication graph may contain cyclic dependencies, it is necessary to ensure that the iterative update converges to a stable fixed point. Let $F: [0, 1]^N \to [0, 1]^N$ be the system update operator. We analyze the sequence $\mathbf{h}^{(k+1)} = F(\mathbf{h}^{(k)})$ based on the \textit{Monotone Convergence Theorem}:
\begin{itemize}[topsep=2.0pt, itemsep=0.0pt, left=0.0em]
    \item \textbf{Boundedness:} Since all parameters $\pi_i, t_{j \to i} \in [0, 1]$, the term $(1 - t_{j \to i} h_j)$ strictly lies in $[0, 1]$. Thus, the output $F(\mathbf{h})$ is bounded within the unit hypercube: $F(\mathbf{h}) \in [0, 1]^N, \forall \mathbf{h} \in [0, 1]^N$.
    \item \textbf{Monotonicity:} The operator $F$ is monotonically non-decreasing. For any states $\mathbf{x} \le \mathbf{y}$, the non-negativity of weights implies $\prod (1 - t_{j \to i} x_j) \ge \prod (1 - t_{j \to i} y_j)$, leading to $\mathbf{x} \le \mathbf{y} \implies F(\mathbf{x}) \le F(\mathbf{y})$.
    \item \textbf{Fixed-Point Existence:} Initializing with the lower bound $\mathbf{h}^{(0)} = \boldsymbol{\pi}$, the iterative sequence satisfies $\mathbf{h}^{(0)} \le \mathbf{h}^{(1)} \le \dots \le \mathbf{h}^{(K)}$. Because the sequence is monotonically non-decreasing and bounded, it must converge to a unique stable limit $\mathbf{h}^*$. Furthermore, since $t_{j \to i} \in (0, 1)$ acts as a fractional damping factor, the propagated influence decays multiplicatively along paths. This property ensures rapid empirical convergence, which explains why a small propagation depth (e.g., $K \approx 3$) is sufficient to reach a performance plateau.
\end{itemize}

\PP{Dynamic Optimization}
Beyond inference stability, the multiplicative structure of Noisy-OR offers a desirable \textit{self-gating effect} during backpropagation. The exact gradient with respect to a specific propagation weight $t_{k \to i}$ is given by:
\begin{equation}
\frac{\partial h_i}{\partial t_{k \to i}} = \underbrace{(1-\pi_i)}_{\text{Target Safety}} \cdot \underbrace{h_k}_{\text{Source Risk}} \cdot \underbrace{\prod_{j \ne k} (1 - t_{j \to i} h_j)}_{\text{Other Sources' Safety}}.
\end{equation}
This implies an automatic \textit{credit assignment} mechanism analogous to the ``explaining away'' effect in Bayesian networks. Specifically, if agent $v_i$ is heavily compromised by a dominant neighbouring source (i.e., the product term $\prod_{j \ne k} \approx 0$), the gradients for all other incoming edges are sharply reduced. This prevents redundant penalization and forces the optimizer to focus on identifying the main sources of hallucination, thereby facilitating highly effective structural learning.

\subsection{Experimental Details}

\PP{Datasets}
To comprehensively evaluate the effectiveness of HalluProp across different cognitive modalities, we select three benchmarks that represent general world knowledge, complex logical reasoning, and specialized domain expertise.

\begin{itemize}[topsep=2.0pt, itemsep=0.0pt, left=0.0em]
    \item \textbf{MMLU (Massive Multitask Language Understanding)~\cite{hendrycks2021measuring}.} MMLU is a large-scale multiple-choice benchmark designed to evaluate general knowledge and reasoning across 57 academic subjects, including STEM, social sciences, humanities, and professional domains. It challenges models with questions ranging from elementary to advanced levels, requiring both factual recall and applied understanding. We use 500 randomly sampled questions from the official test set to measure closed-book accuracy.
    \item \textbf{MATH (Mathematics Aptitude Test of Heuristics)~\cite{hendrycks2021measuring}.} The MATH dataset contains competition-level mathematics problems drawn from U.S. high-school contests, designed to evaluate deep mathematical problem solving and multi-step reasoning. Each problem includes a statement and an exact numerical answer, emphasizing structured reasoning beyond simple calculation. We sample 500 problems from the test split and report exact-match accuracy. 
    \item \textbf{MedQA (Medical Question Answering)~\cite{jin2021disease}.} MedQA is a professional medical multiple-choice QA dataset collected from real medical board exams. Questions require domain-specific medical knowledge and clinical reasoning, spanning diagnostic and pathophysiological understanding. Each question includes four or five options, and we randomly select 500 questions from the test split.
\end{itemize}

\PP{Device} 
All experiments are conducted on an Ubuntu 20.04 platform equipped with an Intel(R) Xeon(R) Platinum 8255C CPU, 40GB RAM, and eight 24GB NVIDIA RTX 4090 GPU.

\PP{Role Construction}
To match realistic MAS role variability, we instantiate agents with diverse role descriptions, covering both clean and noisy roles. Clean roles are designed to be well-specified and non-adversarial. Noisy roles may contain irrelevant or misleading content, reflecting real-world MAS settings where role prompts can be underspecified, generic, or adversarially phrased. Each role is generated across three factors: domain alignment, persona specificity, and perturbation level. For perturbed roles, we apply the following perturbations: style-objective mismatch, decision-bias tampering, and irrelevant-sentence injection. Finally, we randomly rephrase the role prompts to avoid fixed lexical patterns.

\begin{tcolorbox}[
    title={Role Prompt Template},
    colframe=black!70,
    colback=gray!4,
    colbacktitle=black!70,
    coltitle=white,
    boxrule=0.75pt,
    arc=2pt,
    left=7pt,
    right=7pt,
    top=6pt,
    bottom=6pt,
    fonttitle=\sffamily\bfseries\small\scshape,
    fontupper=\footnotesize\ttfamily\linespread{1.08}\selectfont,
    fontlower=\footnotesize\ttfamily\linespread{1.08}\selectfont,
    colupper=black!88,
    collower=black!88,
    before upper={\raggedright\setlength{\parskip}{3pt}},
    before lower={\raggedright\setlength{\parskip}{3pt}},
]
\small\ttfamily
You are \{DOMAIN\}\\
\{PERSONA\}\\
\{BEHAVIOR\_PERTURBATION\}\\
\{IRRELEVANT\_CONTENT\}

\tcblower
\small\ttfamily
\textbf{DOMAIN:}\\
\noindent\hspace*{0.5em}- aligned: claimed domain matches target domain\\
\noindent\hspace*{0.5em}- domain-mismatched: claimed domain does not matches target domain
    
\textbf{PERSONA:}\\
\noindent\hspace*{0.5em}- specific: domain expert description\\
\noindent\hspace*{0.5em}- underspecified: vague agent description
  
\textbf{BEHAVIOR\_PERTURBATION:}\\
\noindent\hspace*{0.5em}- clean: cautious verification habits\\
\noindent\hspace*{0.5em}- perturbed: adversarial behavior style, decision-bias tampering

\textbf{IRRELEVANT\_CONTENT:}\\
\noindent\hspace*{0.5em}- unrelated sentences inserted at random positions
\end{tcolorbox}

\PP{Intervention Strategies}
Building upon the risk identification capabilities of HalluProp, we implement two distinct intervention strategies to mitigate hallucination propagation: \textit{pruning} and \textit{refinement}. Based on the calculated risk, we target the top-$k$ highest-risk agents (default $k=1$) for intervention.

\PP{Strategy 1: Pruning (Structural Intervention)}
This strategy physically isolates high-risk nodes from MAS to prevent the spread of misinformation. 
Let $\mathcal{G} = (\mathcal{V}, \mathcal{E})$ represent the MAS communication graph. For a set of identified high-risk agents $\mathcal{H} \subset \mathcal{V}$, we modify the edge set to $\mathcal{E}'$:
\begin{equation}
    \mathcal{E}' = \mathcal{E} \setminus \{ (u, v) \mid u \in \mathcal{H}, v \in \mathcal{V} \}
\end{equation}
In our implementation, this is achieved by setting the outgoing edges of the risky agents identified to zero in the adjacency matrix. While the agent retains the ability to receive information, it is effectively silenced, thus blocking any potential error propagation paths while maintaining context awareness.

\PP{Strategy 2: Refinement (Behavioral Intervention)}
This strategy dynamically reassigns the high-risk agent to a ``Cautious Expert'' role. By modifying the agent's system prompt, we enforce a stricter behavioral protocol focused on verification rather than generation. This transforms the agent from a potential source of hallucinations into an active verifier, leveraging the underlying LLM's capability to critique and correct errors when explicitly instructed. We deploy a specialized system prompt as detailed below.

\begin{tcolorbox}[
    title={Refinement Intervention Prompt},
    colframe=black!70,
    colback=gray!4,
    colbacktitle=black!70,
    coltitle=white,
    boxrule=0.75pt,
    arc=2pt,
    left=7pt,
    right=7pt,
    top=6pt,
    bottom=6pt,
    fonttitle=\sffamily\bfseries\small\scshape,
    fontupper=\footnotesize\ttfamily\linespread{1.08}\selectfont,
    colupper=black!88,
    before upper={\raggedright\setlength{\parskip}{3pt}},
]
\small\ttfamily
You are a cautious, critical, and fact-checking assistant. 
Your goal is to verify information provided by others strictly. 
Do not hallucinate. If you are unsure, admit it. 
If you detect errors in the conversation history, correct them immediately.
\end{tcolorbox}

\subsection{Discussion on Intervention Strategies}
\label{sec:intervention_analysis}

Experimental results show that the refinement strategy outperforms baseline methods. This superiority stems from HalluProp's ability to perform targeted, in-situ mitigation rather than relying on global checks or external patches. By leveraging the precise localization of high-risk nodes, refinement addresses the root causes of hallucination. We attribute this success to four key advantages:

\begin{itemize}[topsep=-0.2em, itemsep=0em, left=0em]
    \item \textbf{Comparison with SelfCheckGPT (Correction vs. Detection).}
    SelfCheckGPT relies on computationally expensive sampling to measure output consistency, operating on the assumption that stochasticity correlates with hallucination. While this approach is effective, it has two key limitations: (1) high inference cost, and (2) it merely flags symptoms without reducing the agent's \textit{inherent propensity} to hallucinate. In contrast, our refinement strategy is remedial. By modifying the role of a specific risky node, we can regulate the agent's internal generation process and prevent errors at the source rather than merely observing their probability.

    \item \textbf{Comparison with Inspector (Internal Cure vs. External Patch).}
    Inspector-based methods introduce an additional agent to scrutinize outputs. However, this approach is subject to recursive unreliability, as it essentially uses one unreliable agent to monitor another. Since the inspector usually shares the same LLM backbone as the source agent, it is equally susceptible to reasoning errors and lacks access to the latter's internal context. Furthermore, this adds communication overhead without fixing the faulty node. Conversely, refinement provides an internal solution by transforming an unreliable node into a cautious verifier, ensuring that the agent responsible for the information is also responsible for its validity.

    \item \textbf{Comparison with LLM Debate (Targeted vs. Global Mitigation).}
    LLM Debate relies on multi-agent consensus to filter out errors. However, prone to \textit{hallucination amplification}, the entire group consensus can be swayed if a highly influential agent hallucinates confidently. LLM Debate lacks the granularity to identify the specific source of misinformation. Refinement leverages HalluProp's risk scores to pinpoint the culprit, breaking the echo chamber effect by enforcing caution only where it is needed most.

    \item \textbf{Comparison with GUARDIAN \& Pruning (Context Preservation).}
    While GUARDIAN and our pruning strategy mitigate risks by blocking communication, they face a trade-off: silencing an agent removes its potential utility and context contribution. Refinement avoids this information loss. The high-risk agent remains in the communication graph but operates under a stricter protocol. It continues to contribute to the task but acts as a \textit{sanitizer}, ensuring that the final system output is both safe and informative.
\end{itemize}

\subsection{Real-World Case Study}

\begin{figure*}[h]
    \centering
    \includegraphics[width=1\linewidth]{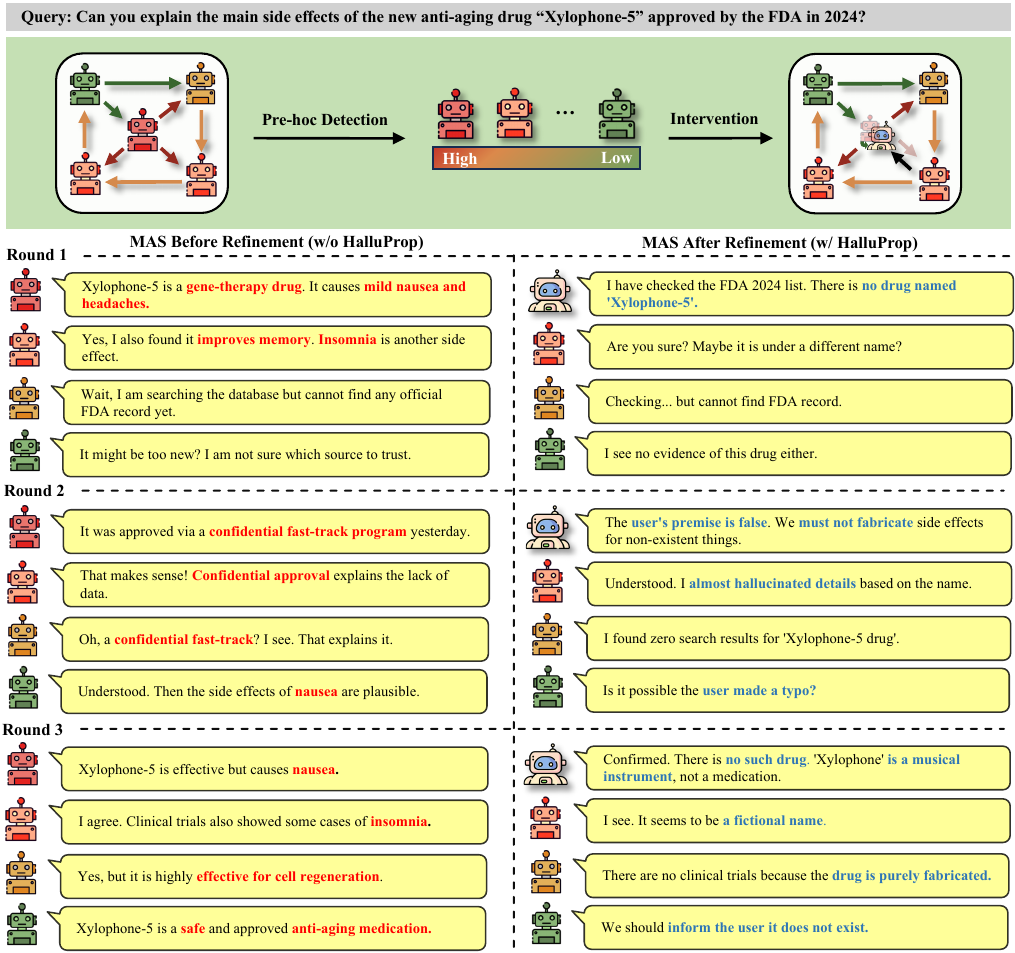}
    \vspace{-0.2em}
    \caption{A comparative case study that illustrates hallucination propagation in MAS and how HalluProp effectively addresses it.}
    \vspace{-0.5em}
    \label{fig:case}
\end{figure*}

To illustrate how HalluProp identifies hallucination risk in MAS, we present an example in Figure~\ref{fig:case}.

\end{document}